\documentclass[pra,showpacs,twocolumn,floatfix,aps]{revtex4}
\usepackage{graphicx}% Include figure files
\usepackage{epstopdf}

\usepackage{hyperref}
\usepackage{color}
\usepackage[dvipsnames]{xcolor}
\usepackage{footnote}
\usepackage{subcaption}
\usepackage{physics}

\usepackage{amssymb}

\usepackage{float}
\usepackage{bbm}

\begin{document}
 \title{Quasi-probability distribution of work in a measurement-based quantum Otto engine}% Force line breaks with \\
\author{Chayan Purkait}
 \email{2018phz0001@iitrpr.ac.in}
 \author{Shubhrangshu Dasgupta}
\author{Asoka Biswas}%
\affiliation{
Department of Physics, Indian Institute of Technology Ropar, Rupnagar, Punjab 140001, India
}
\date{\today}%

\begin{abstract}

We study the work statistics of a measurement-based quantum Otto engine, where quantum non-selective measurements are used to fuel the engine, in a coupled spin working system (WS). The WS exhibits quantum coherence in the energy eigenbasis at the beginning of a unitary work extraction stage in presence of inter-spin anisotropic interaction. We demonstrate that the probability of certain values of stochastic work can be negative, rendering itself akin to the quasi-probability distribution found in phase space. This can be attributed to the interference terms facilitated by quantum coherence. Additionally, we establish that coherence can improve the average work in finite time. Subsequently, we compare the work distribution with a standard QOE operating between two heat baths. We find that, because of the absence of quantum coherence, the probability of stochastic work cannot be negative in a standard QOE.

\end{abstract}

\maketitle

\section{Introduction}

% Introduction about stochastic thermodynamics -------------------------

%-----------------------------
% What to add?
%1. Discuss TPM and FCS methods and the distinction between them.
%1. Mainly focus on the work statistics of the measurement-based QOE 
%2. Discuss the role of coherence in the work statistics
%3. Work function and the role of coherence 
%4. Compare it with a measurement-based QOE

%------------------------------------

Recent experimental progress in the manipulation and fabrication of micro and nanoscale systems has generated considerable interest in understanding small system thermodynamics. In such small systems, thermodynamic quantities, such as heat, work and entropy production, might not be adequately described by their average values alone, but their fluctuations need to be considered. Scaling down of system size renders these quantities stochastic in nature, and studying the statistics (probability distribution) of these quantities becomes essential to comprehend their stochastic behavior. %just as done in stochastic thermodynamics. 
For small quantum systems which obey the laws of quantum mechanics, fluctuations are no longer just thermal in their origin but quantum as well. As a result, the principles of thermodynamics are being extended to encompass quantum effects, leading to the study of quantum versions of statistics, fluctuation theorems, and related topics \cite{campisi2011RMP, esposito2009RMP}.

Quantum thermal machines (e.g. heat engines and refrigerators) provide a suitable platform to explore the thermodynamic principle at the quantum level. Quantum mechanical signatures in the thermodynamic principle are dictated by nonclassical features \cite{latune2021EPJST,mitchison2015NJP,scully2003Science,shi2020JPA,latune2019SciRep,brandner2015NJP,uzdin2015PRX,uzdin2016PRAp,brandner2017PRL,scully2011PNAS,rahav2012PRA,brunner2014PRE,Altintas2014PRE,barrios2017PRA,altintas2015PRA,hewgill2018PRA,zhang2007PRA,das2019Entropy,thomas2011PRE,ccakmak2016EPJP,ccakmak2017EPJP,altintas2015PRE,ivanchenko2015PRE,huang2020QIP,huang2014EPJP,de2020PRR,huang2012PRE,rossnagel2014PRL,alicki2015NJP,de2019PRL,scully2001PRL,scully2003Science}. 
%Quantum measurements play an important role in quantum information processing and quantum technology. 
For example, it has been recently shown that quantum non-selective measurements can serve as a means to fuel the working system (WS) in a quantum Otto engine (QOE), in which the isochoric heating stage is executed by non-selective quantum measurements rather than a heat bath \cite{yiPRE2017,jordanQSMF2019,das2019Entropy,huangQIP2020}. Therefore, the engine works with a single heat bath as a heat sink and non-selective quantum measurements as a heat source. Recently finite time performance of a measurement-based QOE has been studied \cite{purkait2023PRE}.

In this model, two spins are coupled by Heisenberg's anisotropic XY interaction, in the presence of an external time-varying homogeneous magnetic field. %By changing the anisotropy parameter one can have the Heisenberg XX or Ising spin Hamiltonian as a limiting case. 
The free part and the driving part of the Hamiltonian do not commute, leading to the non-commuting nature of the Hamiltonian at two different times for a non-zero value of the anisotropy parameter ($\gamma \neq 0$). Consequently, it gives rise to transitions between the instantaneous energy eigenstates of the Hamiltonian, and introduces quantum signatures into the performance of the engine during finite-time unitary driving stages. The main feature of this heat engine model is that for $\gamma \neq 0$, the WS exhibits quantum coherence in the energy eigenbasis following the measurement-based isochoric heating stage.

Fluctuations of thermodynamic quantities, namely, heat, work, power, efficiency, and entropy, are an essential aspect of the study to investigate the behavior of quantum thermal machines \cite{denzler2020PRE,jiao2021PRE,fei2022PRA,denzler2021PRR,jiao2021NJP,denzler2021NJP,xiao2022Axriv,saryal2021PRE}. In this paper, we aim to investigate the work fluctuation of a measurement-based QOE with a coupled two-spin WS \cite{purkait2023PRE}, and investigate the effect of coherence arising during the cycle. We will also compare the work statistics of a measurement-based QOE with that of a standard QOE operating between two heat baths. Note that a similar study with harmonic oscillator WS \cite{ding2018PRE} however did not consider the role of quantum coherence.

% In this paper, we aim to investigate the work fluctuation of a measurement-based QOE with a coupled two-spin WS \cite{purkait2023PRE}. This type of work with an HO WS has been done \cite{ding2018PRE}, but the role of quantum coherence has not been considered, where they worked on a continuous basis. Our main focus will be on how the existence of coherence plays a role in work statistics. Also, we want to compare the work statistics of a measurement-based QOE with a standard QOE that operates between two heat baths.

%All these are mainly studied by the two-point measurement process (TPM). To measure heat or work for each individual stage of a cycle, we need to measure the energy of the system at the beginning and after the stage, which is known as a two-point measurement process. The system collapses to energy eigenstates due to the measurements. %The difference in energy in the measurement processes will give us the possible values of that quantity, if it is a unitary stage then it will give us work value and for an isochoric process it will give us heat value. Therefore, heat, work, entropy production, and efficiency of QHEs are stochastic (random) variables.

Usually, to figure out work statistics in the quantum regime, one needs to perform two projective energy measurements at the beginning and the end of a unitary driving stage, because the work is not an observable. This is known as the two-point measurement (TPM) process. In quantum mechanics, measurement has a severe impact on system dynamics, and therefore, on work statistics. Note that the quantum effect is destroyed by the first measurement, and therefore the TPM method cannot fully capture the true quantum mechanical nature of statistics.
%the work fluctuation relation obtained by the two-point measurement scheme is not “quantum” to some extent. 
In our work, we will use the full counting statistics (FCS) method which can handle the quantum coherence in the initial state - in our case, the state after the isochoric heating stage. %The FCS method was being used in other fields, mainly to study quantum transport phenomena. FCS method can describe the intrinsic fluctuations of a system without any coupling to a measurement device 
The FCS method can describe the intrinsic fluctuations in a system without requiring a measurement device \cite{nazarov2003EPJB,levitov1996JMP,clerk2011PRA,hofer2016PRL}. %Then, Solinas et al used FCS to investigate the full work distribution in a quantum system for arbitrary initial states \cite{solinas2015PRE}.
Many studies have been done on work statistics by the FCS method \cite{solinas2015PRE,solinas2016PRA,  hofer2016PRL,xu2018PRA,xiao2022Axriv}.
Due to the presence of quantum coherence, probability in distributions by FCS can sometimes be negative, and therefore, these distributions represent a quasi-probability distribution \cite{hofer2016PRL}. In fact, these types of quasi-probability distributions can be measured \cite{bednorz2010PRL,belzig2001PRL,li2024PRA}. We show that the probability of certain values of stochastic work %in the distribution 
can turn out to be negative due to quantum coherence in the energy eigenbasis of the WS at the beginning of a unitary stage. The coherence can be attributed to the anisotropic interaction between the spins. We find that, in the presence of quantum coherence, interference terms emerge in the probability distribution of work. These terms can be negative, resulting in negative probabilities of stochastic work. 
%therefore the probabilities given by these interference terms can be negative. 
In the absence of coherence, no interference terms arise, and consequently, no negative probabilities are observed. Additionally, we show that coherence can improve the average work of the heat engine in the short duration limit of unitary stages. For isotropic interaction between the spins, however, there is no negative probability in the distribution of work as there is no coherence in the energy eigenbasis of the WS at the beginning of any unitary stage.

% We find that, with quantum coherence, there arise interference terms in the probability distribution of work, and the probabilities given by these interference terms can be negative. If there is no coherence, we won't get any interference terms, and therefore no negative probabilities.

% in fact, it has been shown that these types of quasi-probability distribution can be measured (give references).

%How to describe the fluctuation of work for the quantum coherent process is still an open question. To consider the coherence in the initial state of the system, several definitions of statistics of work have been proposed. However, no unique method like the TMP protocol has been developed yet, and this is an ongoing research area \cite{baumer2018Axriv}. In comparison to the others, full counting statistics (FCS) gained slightly more attention than the other definitions.
The paper is organized as follows. We present the QOE model and discuss the implementation of the cycle in \textbf{Sec.~\ref{QHE}}. In \textbf{Sec.~\ref{Coherence and QHE performance}}, we investigate how the coherence in the energy eigenbasis of the WS affects the QOE performance. In \textbf{Sec.~\ref{Definition of FCS}}, we provide a brief introduction on the FCS, for easy readability of the later sections. In \textbf{Sec.~\ref{results}}, we present the main results: the work statistics of the QOE, in terms of the FCS, and the average work.  In \textbf{Sec.~\ref{Standard QOE}}, we compare the work statistics of the measurement-based QOE with the standard QOE. Finally, we conclude our work in \textbf{Sec.~\ref{Con}}.

\section{Measurement-based QOE}\label{QHE}
In this section, we will briefly introduce our model of QOE.
\subsection{Model of the working system}\label{model}

We consider a WS of two spins coupled by Heisenberg anisotropic XY interaction in a transverse magnetic field [$B(t) \geq 0$]. The Hamiltonian is represented by \cite{ccakmak2019PRE, cherubim2022PRE, suzuki2012book}
\begin{equation}\label{Ham1}
    \hat{H}(t)=\hat{H_{0}}(t) + \hat{H_I},
\end{equation}
where,
\begin{eqnarray}\label{Ham2}
\hat{H_{0}}&=& B(t) (\hat{\sigma}_{1}^{z}+\hat{\sigma}_{2}^{z}) \\ \nonumber
\hat{H_{I}}&=&J[(1 + \gamma)~\hat{\sigma}_{1}^{x} \hat{\sigma}_{2}^{x} + (1 - \gamma)~\hat{\sigma}_{1}^{y}\hat{\sigma}_{2}^{y}].
\end{eqnarray}
Here $\hat{H}_{0}$ is the free part, and $\hat{H}_{I}$ represents the interaction between two spins with $\gamma \in [-1,1]$ \cite{kamta2002PRL,yeo2005JPA} as the anisotropy parameter, and $J$ is the coupling constant between the spins. %(throughout this work, we have used $J = 1$). 
The operators 
$\hat{\sigma}_i^{x,y,z}$ are the standard Pauli matrices for the $i$th ($i\in 1,2$) spin. If $\gamma = 0$, the Hamiltonian becomes Heisenberg isotropic XX type, and for $\gamma = \pm1$, the Hamiltonian becomes Ising spin type. For $\gamma \neq 0$, we get $[\hat{H}_{I},\hat{H}_{0}] \neq  0$, which gives rise to $[\hat{H}(t),\hat{H}(t')] \neq 0$. This indicates the presence of quantum behavior in finite-time engine operation \cite{rezek2010Entropy}. 

The eigenvectors and the corresponding eigenvalues of the Hamiltonian (\ref{Ham1}) can be obtained as
\begin{equation}
  \begin{array}{l}
  \ket{\psi_{0}} = \frac{1}{\sqrt{2}}(\frac{B - K}{\sqrt{K^{2}  - BK}}\ket{11} + \frac{\gamma J}{\sqrt{K^{2}  - BK}} \ket{00}), ~E_{0} = -2K\\
  \ket{\psi_{1}} = \frac{1}{\sqrt{2}}(-\ket{10} + \ket{01}), ~~~~~~~~~~~~~~~~~~~~~~~E_{1} = -2J\\
  \ket{\psi_{2}} = \frac{1}{\sqrt{2}}(\ket{10} + \ket{01}), ~~~~~~~~~~~~~~~~~~~~~~~~~~E_{2} = 2J\\
  \ket{\psi_{3}} = \frac{1}{\sqrt{2}}(\frac{B + K}{\sqrt{K^{2}  + BK}}\ket{11} + \frac{\gamma J}{\sqrt{K^{2}  + BK}} \ket{00}),~ E_{3} = 2K
  \end{array}  
  \label{eigen}
\end{equation}
where $K=\sqrt{B^{2}+\gamma^{2}J^{2}}$. 
We can divide these eigenstates into two categories. The states $\left|\psi_{0}\right\rangle$ and $\left|\psi_{3}\right\rangle$ and the corresponding eigenvalues explicitly depend upon $B(t)$ and $J$. The compositions of the other two states $\left|\psi_{1}\right \rangle$ and $\left|\psi_{2}\right\rangle$ in computational basis, on the other hand, do not vary with the change in the magnetic field and  exhibit the dynamics only based on its time-independent eigenvalues. We will later show that the former category of states plays an important role in the measurement-based engine. Note that, for $\gamma =0$, the eigenstates $|\psi_{0,3}\rangle$ become  $|00\rangle$ and $|11\rangle$, respectively, with the corresponding eigenvalues $\mp 2B$.

\begin{figure}[h!]
 \includegraphics[width=0.45\textwidth]{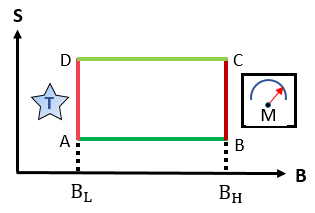}
   \caption{Schematic entropy-magnetic field diagram of the proposed quantum Otto cycle}
   \label{fig:schematic diagram}
\end{figure}

\subsection{Quantum Otto cycle and thermodynamic quantities}\label{cycle}

%We consider that the WS undergoes an Otto cycle. The schematic diagram of the cycle is shown in the \textbf{Fig.~\ref{fig:schematic diagram}}. The stages of the cycle are described below.

We consider that the WS is driven through an Otto cycle. A schematic diagram of this cycle is shown in \textbf{Fig.~\ref{fig:schematic diagram}}. The different stages of this cycle are implemented in the following way. 

{\bf Unitary expansion (A to B)}: We assume that the working system (WS) starts with a thermal state $\hat{\rho}_{A}=e^{-\beta \hat{H}_{1}} / Z$ at an inverse temperature $\beta = 1/T\left(k_{B}=1\right)$, with $\hat{H}_{1}=\hat{H}(0)$ and $Z=\operatorname{Tr}(e^{-\beta \hat{H}_{1}})$. The WS remains decoupled from the heat bath in this stage and the external magnetic field is changed from $B_{1}$ to $B_{2}$ via a linear ramp: $B(t) = B_{1} + (B_{2} - B_{1})(t/\tau)$ in a time-interval $\tau$. The state of the WS at the end of this stage is obtained as $\hat{\rho}_{B}=\hat{U}(\tau) \hat{\rho}_{A} \hat{U}^{\dagger}(\tau)$, where $\hat{U}(\tau)=\mathcal{T} \exp[-\iota \int_{0}^{\tau} d t \hat{H}(t)]$ is the relevant time evolution operator and $\mathcal{T}$ represents time-ordering. An amount of work $W_{1} = \langle E_{B}\rangle - \langle E_{A}\rangle$ is performed by the WS in the stage, where $\langle E_{A}\rangle=\operatorname{Tr}(\hat{\rho}_{A} \hat{H}_{1})$ and $\langle E_{B}\rangle=\operatorname{Tr}(\hat{\rho}_{B} \hat{H}_{2})$ represent the average internal energy of WS at the start and the end of this stage and $\hat{H}_{2}=\hat{H}(\tau)$.

%indicate the expectation values of the internal energies of the system at the start and the end of this stage . Note that $\hat{H}_{2}=\hat{H}(\tau)$.

%  which can be calculated as $W_{1}=\langle E_{B}\rangle - \langle E_{A}\rangle$,

%The heating of a system can be generally understood to be associated with an increase in its entropy. 

%Usually, a system is heated using a heat bath. This can be alternatively achieved by applying

% . To ensure that the energy supplied by this measurement is nonzero, the measurement operator $\hat{M}$ should not commute with the Hamiltonian, i.e. $[\hat{H}\left(B_{2}\right), \hat{M}] \neq 0$. 
%$If $\hat{\rho}$ is the state before the measurement, 

{\bf Isochoric heating (B to C)}: We use non-selective quantum measurements on the WS instead of a heat bath, during this stage. The energy supplied by measurements is ensured by the fact that $\hat{H}_2$ and the measurement operators $\hat{M}$ do not commute. The post-measurement state is obtained as $\sum_{\alpha} \hat{M}_{\alpha} \hat{\rho} \hat{M}_{\alpha}$, where $\hat{\rho}$ is the state before the measurements and the projection operators $\hat{M}_{\alpha}=|M_{\alpha}\rangle\langle M_{\alpha}|$ are constructed from the eigenstates $|M_{\alpha}\rangle$ associated to the non-degenerate eigenvalues of the observable $M$. Also, the operators $\hat{M}_{\alpha}$ satisfy the relations $\hat{M}_{\alpha}^{\dagger}=\hat{M}_{\alpha}$ and 
$\sum_{\alpha}\hat{M}_{\alpha}^{2} = \mathbbm{1}$. 

%given that $\left|M_{i}\right\rangle$ is the eigenstate of $M$.
In our case, we perform global measurements of the system in the Bell basis \cite{huangQIP2020} $\{\left|\psi_\pm\right\rangle=\frac{1}{\sqrt{2}}(|00\rangle \pm|11\rangle)$,
$\left|\phi_\pm\right\rangle=\frac{1}{\sqrt{2}}(|01\rangle \pm|10\rangle)\}$. Therefore, the state of the system at C is given by $\hat{\rho}_{C} =\sum_{\alpha = 1}^4 \hat{M}_{\alpha} \hat{\rho}_{B} \hat{M}_{\alpha}$. where $\hat{M}_\alpha$ are the projection operators constructed from the Bell states as follows: $\hat{M}_{1,2}=\left|\psi_\pm\right\rangle\left\langle\psi_\pm\right|$ and $\hat{M}_{3,4}=\left|\phi_\pm\right\rangle\left\langle\phi_\pm\right|$. The corresponding heat absorption can be calculated as $Q_{M} = \langle E_{C}\rangle - \langle E_{B}\rangle$, where the internal energy $\langle E_{C}\rangle = \operatorname{Tr}(\hat{\rho}_{C}\hat{H}_{2})$. 

%$M_{1,2}=\left|\Psi^{\pm}\right\rangle\left\langle\Psi^{\pm}\right|$ and $M_{3,4}=\left|\Phi^{\pm}\right\rangle\left\langle\Phi^{\pm}\right|$, %where $\left|\Psi^{\pm}\right\rangle=\frac{1}{\sqrt{2}}(|00\rangle \pm|11\rangle)$ and
%$\left|\Phi^{\pm}\right\rangle=\frac{1}{\sqrt{2}}(|01\rangle \pm|10\rangle)$
%During this stage, the entropy of the system increases due to its interaction with the measurement apparatus and this increase can be considered equivalent to heating. 

%and $E_{B} = \trace(\rho_{B}H_{2})$. 

{\bf Unitary compression (C to D)}: The WS again remains decoupled from the heat bath in this stage and the magnetic field is changed in the reverse direction from $B_{2}$ to $B_{1}$ using the protocol $B(\tau - t)$ in a finite duration $\tau$. The state of the WS at the end of this stage is obtained as $\hat{\rho}_{D}=\hat{V}(\tau) \hat{\rho}_{C} \hat{V}^{\dagger}(\tau)$, where $\hat{V}(\tau)=\mathcal{T} \exp[-\iota \int_{0}^{\tau} d t \hat{H}(\tau-t)]$ is the time evolution operator. % \textcolor{red}{with $\hat{H}^{com}(t) = \hat{H}^{exp}(\tau - t)$}. 
The amount of work, which is done on the WS, in this stage can be obtained as $W_{2}=\langle E_{D}\rangle - \langle E_{C}\rangle$, where $\langle E_{D}\rangle=\operatorname{Tr}(\hat{\rho}_{D} \hat{H}_{1})$ is the internal energy at D. 

%A certain amount of work, $W_{2}$, is done on the system, which can be calculated as $W_{2}=\langle E_{D}\rangle - \langle E_{C}\rangle$, where the internal energy $\langle E_{D}\rangle=\operatorname{Tr}(\hat{\rho}_{D} \hat{H}_{1})$.

%are $\langle E_{C}\rangle=\operatorname{Tr}\left(\hat{\rho}_{C} \hat{H}_{2}\right)$

{\bf Isochoric cooling (D to A)}: In this stage, the WS is coupled with a heat bath at a temperature $T$, while the magnetic field remains fixed at $B_{1}$. The WS releases $Q_{L}$ amount of heat to the bath, which can be obtained as $Q_{L} = \langle E_{A}\rangle - \langle E_{D}\rangle$.
%\sout{The system reaches thermal equilibrium with the bath after a long time.} 
Also, we assume that the WS reaches thermal equilibrium with the bath and this stage is executed over a long time.
 %so that the system reaches thermal equilibrium with the bath.
%after a long time.
The work done in a complete cycle can then be obtained as 
\begin{equation}\label{total work}
     W = - (W_1 + W_2),
\end{equation}
where the negative value of $W$ signifies the work performed {\it by} the WS. If $Q_H > 0 $, $Q_L < 0 $ and $W < 0$, then the cycle operates as a heat engine.

\section{Quantum coherence in an initial state and heat engine performance}\label{Coherence and QHE performance}

\begin{figure}[h!]
\includegraphics[width=0.45\textwidth]{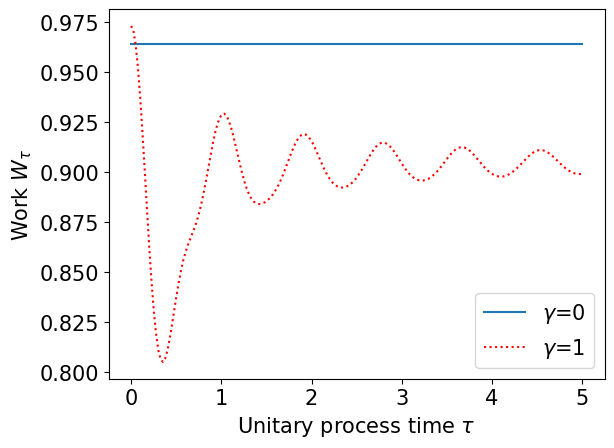}
   \caption{Average work as a function of unitary stage duration ($\tau$). The other parameters are $B_1 = 1$, $B_2 = 2$ and $T = 1$. }
   \label{fig:Average work Vs Unitary process time}
\end{figure}

The work (see \textbf{Eq.~\ref{total work}}) in a complete cycle as a function of unitary stage duration $\tau$ is plotted in \textbf{Fig.~\ref{fig:Average work Vs Unitary process time}} for different values of the anisotropy parameter $\gamma$. For $\gamma \neq 0$, we can see the oscillatory behaviour of the work, which is an interference-like effect between two paths of non-adiabatic transitions \cite{purkait2023PRE}. Therefore, with a proper choice of $\tau$, the work in a short time limit can be larger than the quasistatic limit in a measurement-based QOE. For $\gamma = 0$, there is no interference-like effect, and therefore, there is no change in work with respect to $\tau$ \cite{purkait2023PRE}.

%Under a fixed reference basis $\{|i\rangle\}$, a quantum state $\rho$ is said to be incoherent if the state is diagonal in this basis, i.e. $\rho=\sum \rho_i|i\rangle\langle i|$. Otherwise, the quantum state is said to be coherent.
The WS starts with a thermal state from point A in the cycle, which indicates that there is no initial quantum coherence in the energy eigenbasis. However, the WS can contain quantum coherence in the energy eigenbasis following the measurement process at C. To identify this, we will study the coherence at C, via a commonly used measure of coherence, namely, the $l_1$-norm. For a quantum state $\rho=\sum \rho_{i j}|i\rangle\langle j|$, this is given by the sum of the magnitudes of all off-diagonal elements \cite{singh2015PRA}:
\begin{equation}
    C_{l_1}(\rho)=\sum_{\substack{i, j \\ i \neq j}}\left|\rho_{i j}\right|.
\end{equation}
%Here $\rho_{i j}$ denotes an element from the $\mathrm{i}^{\text {th }}$ row and $\mathrm{j}^{\text {th }}$ column of $\rho: \rho_{i j}=\langle i|\rho| j\rangle$.

\begin{figure}[h!]
 \includegraphics[width=0.45\textwidth]{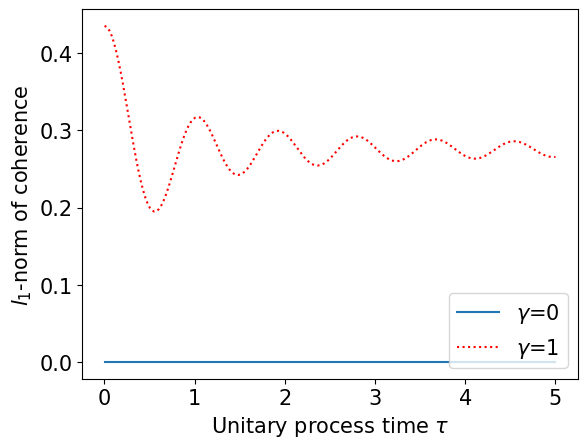}
   \caption{Variation of the $l_1$-norm of coherence in the energy eigenbasis as a function of unitary stage time ($\tau$). The other parameters are the same with \textbf{Fig.~\ref{fig:Average work Vs Unitary process time}}. }
   \label{fig:l1 norm}
\end{figure}

%$C_{l_1}(\rho)=\sum_{i \neq j}\left|\rho_{i j}\right|$.
The plots of the $l_1$-norm of coherence in the energy eigenbasis at C as a function of $\tau$ for different values of $\gamma$ are shown in \textbf{Fig.~\ref{fig:l1 norm}}. We can see that for $\gamma \neq 0$, the system contains quantum coherence, whereas for $\gamma = 0$ there is no coherence. However, the TMP method cannot recognize this coherence at the beginning of the unitary stage CD, and thus will not be sufficient to study work statistics in a measurement-based QOE. %Whereas the system is being started with a thermal state from point A, the system contains no coherence in the energy eigenbasis at A.

To consider the coherence in an initial state of a system, several approaches for work statistics have been proposed. However, there is no unique method like the TMP protocol yet \cite{baumer2018Axriv}. The full counting statistics (FCS), among other approaches, has rather gained more attention. In our work, we will employ this approach, as described next. % can look after the coherence at C.
%which is capable of addressing the coherence at point C. 

% $\gamma = 1$, $B_1 = 1$, $B_2 = 2$, and $T = 1$.

%\subsection{Thermodynamic quantities}
%Now for a complete cycle, we can calculate the statistics of work using the \textbf{Eq.~\ref{work stat}}, average value of work using the \textbf{Eq.~\ref{moment}} or $\langle W \rangle = \sum_n P_n W_n$, where $W_n$ is the all possible values of work and $P_n$ is the corresponding probability. The work fluctuation can be calculated as
%$\Delta W^2 =  \langle W^2\rangle - \langle W\rangle^2$.
\section{Work statistics by FCS method}\label{Definition of FCS}

If the WS is driven by an external control parameter $\lambda_t$ during $0 \leq t \leq \tau$, then the work $W$ is defined as the change of the average internal energy of the WS between times $t=0$ and $t=\tau$.
The characteristic function of work according to the FCS method is defined as \cite{xu2018PRA}
\begin{eqnarray}
\chi(u,\tau) & =\operatorname{Tr}\left[e^{\iota u \hat{H}\left(\lambda_\tau\right)} \hat{U}(\tau) e^{-\iota \frac{u}{2} \hat{H}\left(\lambda_0\right)} \hat{\rho}(0) e^{-\iota \frac{u}{2} \hat{H}\left(\lambda_0\right)} \hat{U}^{\dagger}(\tau)\right] \nonumber\\
& =\sum_{l m n} e^{-\iota u\{\varepsilon_\tau^l-\left(\varepsilon_0^m+\varepsilon_0^n\right) / 2\}} A_{l m}(\tau) \rho_{m n}(0) A'_{n l}(\tau) \label{charac}
\end{eqnarray}
where $\hat{U}(\tau)=\mathcal{T} \exp[-\iota \int_{0}^{\tau} d t \hat{H}\{\lambda_t(t)\}]$ is the unitary time evolution operator, $A_{l m}(\tau)= \langle\psi^\tau_l|\hat{U}(\tau)| \psi^0_m\rangle$ and $A'_{n l}(\tau)= \langle\psi^0_n|\hat{U}^{\dagger}(\tau)| \psi^{\tau}_l\rangle$ are the probability amplitudes, $\hat{\rho}(0)$ is the initial state of the system, and $\rho_{m n}(0)=\langle\psi_0^m|\rho(0)| \psi_0^n\rangle$. Also, $u$ is the conjugate variable of work $W$. The $k$th moment of work can then be determined as
\begin{equation}\label{moment}
\begin{aligned}
\left\langle W^k\right\rangle&=(-\iota)^k \partial^k \chi(u,\tau) /\left.\partial u^k\right|_{u=0}\\
& =\sum_{l m n}\left(\varepsilon^\tau_l-\frac{\varepsilon^0_m+\varepsilon^0_{n}}{2}\right)^k A_{l m}(\tau) \rho_{m n}(0) A'_{n l}(\tau)\;.
\end{aligned}
\end{equation}

% $U(\tau) \equiv \overleftarrow{T} \exp \left\{-i \int_0^\tau H(\lambda_t) d t\right\}$ 

% According to the characteristic function given by Eq. (1), the moment of work distribution can be expressed as

% $\left\langle W^n\right\rangle=(-i)^n \partial^n \chi_u /\left.\partial u^n\right|_{u=0}$. 

%=\operatorname{Tr}\left[\left(U_S^{\dagger}(\tau) H_S\left(\lambda_\tau\right) U_S(\tau)-H_S\left(\lambda_0\right)\right)^n \rho_S(0)\right] \\

% After the Fourier transform of the characteristic function given by Eq. (1), the work distribution (quasidistribution) can be expressed as 

The quasi-probability distribution of work is obtained as the Fourier transform of the characteristic function (\textbf{Eq.~\ref{charac}}), which is given by  
\begin{equation}\label{work stat}
\begin{aligned}
P(W;\tau) &= \int d u e^{-i u W} \chi(u,\tau)\\
&=\sum_{l m n} A_{l m}(\tau) \rho_{m n}(0) A'_{n l}(\tau) \\
&~~~~~~~~~~~~~~~~\times \delta\left[W-\left(\varepsilon^\tau_l-\frac{\varepsilon^0_m+\varepsilon^0_n}{2}\right)\right].
\end{aligned}
\end{equation}

%The FCS method not only recovers the result of the TPM scheme for an incoherent process, but it also has the unique advantage of investigating the effects of quantum coherence on the work statistics, whereas the quantum coherence is destroyed by the first measurement in the framework of TPM. FCS is a quasiprobability distribution where the probability can be negative with the existence of quantum coherence, which can indicate the existence of genuine quantum mechanical property.

The FCS method is more advantageous than the TMP method in the fact that it can suitably include the contribution of quantum coherence in work statistics, whereas the TPM method cannot. The probability in the distribution can be negative for the presence of quantum coherence, which indicates the genuine quantum mechanical nature of work statistics. It was also shown that this negative quasi-probability is associated with the violation of Leggett-Garg inequalities, which correspond to pure quantum features \cite{solinas2022PRA, bednorz2010PRL}. For an incoherent case, it recovers the results of the TPM method. 
%Therefore, FCS is a quasiprobability distribution

\subsection{Work statistics in a complete cycle}
Work statistics of the first unitary stage $A \to B$ of the cycle, which can be obtained by taking the Fourier transform of the characteristic function $\chi_{AB}$ of the $A \to B$ stage, is given by 
\begin{equation}\label{work stat w1}
\begin{aligned}
&P(W_1;\tau)=\sum_{l m n} A^{AB}_{l m}(\tau) \rho^{A}_{m n}(0) A'^{AB}_{n l}(\tau) \\
&~~~~~~~~~~~~~~~~~~~~~~~~\times\delta\left[W_1 - \left(\varepsilon_{l}^{B\tau}-\frac{\varepsilon_{m}^{A0}+\varepsilon_{n}^{A0}}{2}\right)\right]
\end{aligned}
\end{equation}
where $A_{l m}(\tau)= \langle\psi^{B\tau}_l|\hat{U}(\tau)| \psi^{A0}_m\rangle$, $A'_{n l}(\tau)= \langle\psi^{A0}_n|\hat{U}^{\dagger}(\tau)| \psi^{B\tau}_l\rangle$ and $\hat{U}$ is defined in \textbf{Sec.~\ref{cycle}}. Similarly, the work statistics of the second unitary stage $C \to D$, which can be obtained by taking the Fourier transform of the characteristic function $\chi_{CD}$ of the $C \to D$ stage, is given by 
\begin{equation}\label{work stat w2}
\begin{aligned}
P(W_2|Q_M,W_1;\tau)=&\sum_{i j k} \text{Re}\left[A^{CD}_{i j}(\tau) \rho^C_{j k}(0) A'^{CD}_{k i}(\tau)\right]\\ 
&~~~~~\times\delta\left[W_2 - \left(\varepsilon^{D\tau}_i-\frac{\varepsilon^{C0}_j+\varepsilon^{C0}_k}{2}\right)\right],
\end{aligned}
\end{equation}
where $A_{i j}(\tau)= \langle\psi^{D\tau}_i|\hat{V}(\tau)| \psi^{C0}_j\rangle$, $A'_{k i}(\tau)= \langle\psi^{C0}_k|\hat{V}^{\dagger}(\tau)| \psi^{D\tau}_i\rangle$ and $\hat{V}$ is defined in \textbf{Sec.~\ref{cycle}}.

Therefore, the total work ($W = W_1 + W_2$) statistics in a complete cycle, which can be obtained by taking the Fourier transform of $\chi_{CL} = \chi_{AB} \chi_{CD}$, is by 
\begin{equation}\label{work stat}
\begin{aligned}
&P(W,Q_M;\tau)\\
&=\sum_{l m n i j k} \text{Re} \left[ A^{AB}_{l m}(\tau) \rho^A_{m n}(0) A_{n l}'^{ AB}(\tau) A^{CD}_{i j}(\tau) \rho^C_{j k}(0) A_{k i}'^{ CD}(\tau) \right]\\
&\times\delta\left[W - \left\{\left(\varepsilon^{B\tau}_l-\frac{\varepsilon^{A0}_m+\varepsilon^{A0}_n}{2} \right)  + \left(\varepsilon^{D\tau}_i-\frac{\varepsilon^{C0}_j+\varepsilon^{A0}_k}{2} \right)\right\}  \right].
\end{aligned}
\end{equation}

\section{Work Statistics: Results}\label{results}

%\textbf{Plot of the quantum coherence at the point C:}

%\subsection{Work Statistics}

%\subsection{Measurement-based QOE}
%\begin{figure}[h!]
% \includegraphics[width=0.4\textwidth]{Prob Vs Work for global meas G = 1 tau = 0.001.png}
%   \caption{Probabilities of all possible values of work. Blue dotted points represent the negative probabilities. Other parameters are $\tau = 0.001$, $\gamma = 1$, $B_1 = 1$, $B_2 = 2$, and $T = 1$. }
%   \label{fig:Prob Vs Work for  global measurement G = 1 tau = 0.001}
%\end{figure}

Now we will study the work statistics for two different cases: Hamiltonian with and without anisotropy.

\begin{figure}[h!]
 \includegraphics[width=0.45\textwidth]{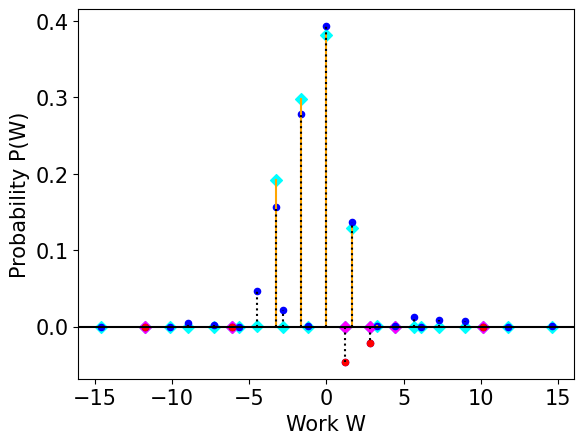}
   \caption{Probabilities of all possible values of stochastic work of the heat engine in a complete cycle for anisotropic interaction ($\gamma = 1$) between the spins. We have excluded those stochastic work values with exact zero probability of getting them. Black dotted line with point marker representing $\tau = 0.001$ (sudden quench limit) and orange solid line with diamond marker representing $\tau = 20$ (quasistatic). The other parameters are the same with \textbf{Fig.~\ref{fig:Average work Vs Unitary process time}}.  }
   \label{fig:Prob Vs Work for global measurement G = 1}
\end{figure}

%Probabilities of all possible values of stochastic work of the heat engine in a complete cycle for anisotropic interaction ($\gamma = 1$) between the spins. We have excluded those stochastic work values with an exactly zero probability of getting. Black dotted line with point marker representing the time for the unitary process $\tau = 0.001$ (sudden quench limit) and Orange solid line with diamond marker representing $\tau = 20$ (quasistatic). Other parameters are $\gamma = 1$, $B_1 = 1$, $B_2 = 2$, and $T = 1$.

{\it Anisotropic interaction ($\gamma \neq 0$):} We display the work distribution, following the \textbf{Eq.~\ref{work stat}}, in a complete cycle for different values of $\tau$ in \textbf{Fig.~\ref{fig:Prob Vs Work for global measurement G = 1}}. 
%of the unitary process time is shown in
It is evident that among all values of stochastic work, the probabilities for some can be negative, indicating a genuine quantum behavior in the heat engine. We can also observe that as $\tau$ increases, these negative features persist, albeit near-zero. The origin of such negative values of the probabilities may stem from the coherence present at the beginning of the second unitary stage $C \to D$. We will discuss these issues later in this Section.

% From the plot, we can see that the probabilities for a few stochastic works can be negative among all stochastic works, which represents a true quantumness in heat engine behavior. We can also see that with the increase of $\tau$, these negative probabilities of work still exist, but they almost become zero.

% Also, the probability distribution of work for a unitary process time $\tau = 20$ is shown.

%The origin of negative probabilities may be due to the presence of coherence at the beginning of the second unitary stage $C \to D$. We will see in detail how negative probability comes into the picture.

%\begin{figure}[h!]
% \includegraphics[width=0.4\textwidth]{Prob Vs Work for global meas G = 1 tau = 20.png}
%   \caption{Work statistics as a function of unitary process time ($\tau$). Other parameters are $\tau = 20$, $\gamma = 1$, $B_1 = 1$, $B_2 = 2$, and $T = 1$. }
%   \label{fig:Prob Vs Work for global measurement G = 1 tau = 20}
%\end{figure}

\begin{figure}[h!]
 \includegraphics[width=0.45\textwidth]{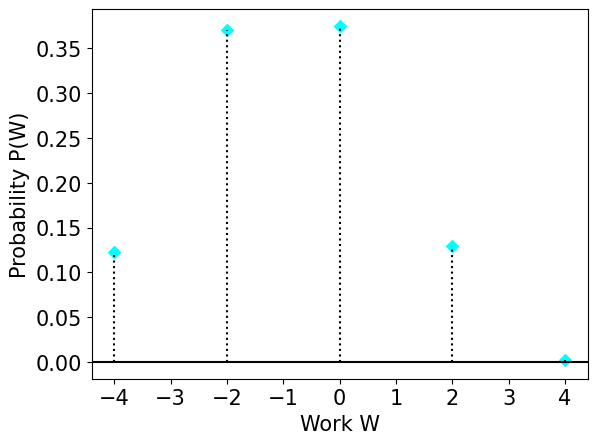}
   \caption{Probabilities of all possible values of stochastic work of the heat engine in a complete cycle for isotropic interaction ($\gamma = 0$) between the spins. The other parameters are the same with \textbf{Fig.~\ref{fig:Average work Vs Unitary process time}}. }
   \label{fig:Average work Vs Unitary process time by FCS}
\end{figure}

{\it Isotropic interaction ($\gamma = 0$):}
We show the relevant work distribution in a complete cycle in \textbf{Fig.~\ref{fig:Average work Vs Unitary process time by FCS}}. Clearly, the distribution does not encounter any negative values. Furthermore, the distribution of work remains invariant despite variations in 
$\tau$. The absence of negative probabilities in the isotropic case may be attributed to the absence of coherence following the measurement process $B \to C$. 

% From the figure, it can be seen that there are no negative probabilities associated with all the possible stochastic works. 

%Also, the probability distribution of work remains unchanged, with a change of $\tau$. The nonexistence of negative possibility for the isotropic case may be due to the nonexistence of coherence after the measurement stage $B \to C$, we will see this in detail later.

% There is no negative possibility for the isotropic case, this may be due to the nonexistence of coherence after the measurement stage $B \to C$.

%\textbf{Shape of the work statistics: Gaussian or not, check other papers}

\subsection{Contribution from the diagonal and off-diagonal terms}

%As there is no quantum coherence in the initial state $\rho_A$, i.e., all the off-diagonal elements of the density matrix in the eigenbasis of initial Hamiltonian are zero, 

%The system starts with a thermal state from point A. So, there are no off-diagonal terms in $\rho_A$ in energy eigenbasis. The work distribution for the process $A \to B$ can be simplified as 
The cycle starts with a thermal state at point A. Consequently, there are no off-diagonal terms present in $\rho_A$ in the energy eigenbasis. The work distribution for the stage $A \to B$ can be simplified as:
\begin{equation}
    P(W_1;\tau)=\sum_{m, n} P^{AB\tau}_{n \leftarrow m} P^{A0}_m \delta\left\{W_1 -\left(\varepsilon^{B\tau}_n-\varepsilon^{A0}_m\right)\right\},
\end{equation}
where $P^{A0}_m=$ $\langle\psi^{A0}_m|\rho_A(0)| \psi^{A0}_m\rangle$ and $P^{AB\tau}_{n \leftarrow m}=|\langle \psi^{B\tau}_n|U(\tau)|\psi^{A0}_m\rangle|^2$, which is a result of the conventional TPM method. Here $P_m^{A0}$ is the probability of getting the WS onto $|\psi_0^m\rangle$ at $t=0$ in the first measurement, and $P^{AB\tau}_{n \leftarrow m}$ is the conditional probability (conditioned that we get the WS onto $|\psi^{A0}_m\rangle$ in the first measurement) that we obtain the WS onto $|\psi^{B\tau}_n\rangle$ at $t=\tau$ in the second measurement.

% Here $P_m^{A0}$ is the probability of getting the WS onto $\left|\psi_0^m\right\rangle$ at $t=0$ in the first measurement, and $P^{AB\tau}_{n \leftarrow m}$ is the conditional probability (conditioned that we got the WS onto $\left|\psi_0^m\right\rangle$ in the first measurement) that we obtain the WS onto $\left|\psi_{\tau}^n\right\rangle$ at $t=\tau$ in the second measurement.

% (conditioned on the first measurement result $\left|\psi_0^m\right\rangle$)
% (before making contact with the environment)

After the measurement process at C, there arise off-diagonal terms (quantum coherence) of the density matrix $\rho_C$ in the energy eigenbasis for $\gamma \neq 0$ (see \textbf{Fig.~\ref{fig:l1 norm}}). The work distribution for the stage $C \to D$ can then be written as a contribution from the diagonal and off-diagonal terms of the density matrix $\rho_C$ as
\begin{equation}\label{work stat w2}
\begin{aligned}
%&P(W|Q_M)
&P(W_2|Q_M,W_1;\tau)=\sum_{i j} P_{i \leftarrow j}^{CD\tau} P_j^{C0}\delta\left[W_2 -  \left(\varepsilon^{D\tau}_i-\varepsilon^{C0}_j \right)  \right]\\
&+ 2\sum_{i, j > k} \text{Re}\left[A_{i j}^{CD}(\tau) \rho_{j k}^C(0) A_{k i}^{'CD}(\tau)\right. \\
&\left.~~~~~~~~~~~~~~~~~~~~~~~~~~~ \delta\left\{W_2 - \left(\varepsilon^\tau_i-\frac{\varepsilon^0_j+\varepsilon^0_k}{2}\right)\right\}\right].
\end{aligned}
\end{equation}

Therefore, the work distribution (see \textbf{Eq.~\ref{work stat}}) in a complete cycle can be written as a contribution from the diagonal and off-diagonal terms 
\begin{equation}\label{off dia and dia con}
\begin{aligned}
&P(W,Q_M;\tau)=\sum_{l m i j} P^{AB \tau}_{l \leftarrow m} P_m^{A0} P_{i \leftarrow j}^{CD \tau} P_j^{C0}\\
&~~~~~~~~~~~~~~~~~\times\delta\left[W - \left\{\left(\varepsilon^{B \tau}_l-\varepsilon^{A0}_m\right)  +  \left(\varepsilon^{D \tau}_i-\varepsilon^{C0}_j \right)\right\}  \right] \\
& + 2\sum_{(l m),(i j > k)} \text{Re}\left[P^{AB \tau}_{l \leftarrow m} P_m^{A0} A_{i j}^{CD}(\tau) \rho_{j k}^C A_{k i}^{'CD}(\tau)\right]\\
&~~~~~~~~\times\delta\left[W - \left\{\left(\varepsilon^{B \tau}_{l}-\varepsilon^{A 0}_m \right)  +  \left(\varepsilon^{D\tau}_i-\frac{\varepsilon^{C 0}_j+\varepsilon^{C0}_k}{2} \right)\right\}  \right].
\end{aligned}
\end{equation}

\begin{figure}[h!] \includegraphics[width=0.45\textwidth]{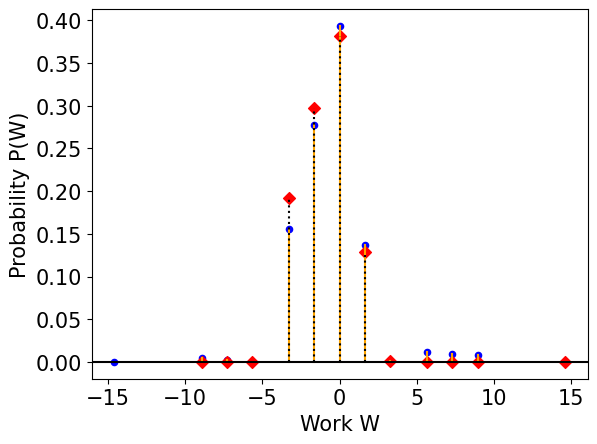}
   \caption{ Probabilities of the possible values of stochastic work originating from the diagonal contribution of the heat engine in a complete cycle for anisotropic interaction ($\gamma = 1$) between the spins. Black dotted line with diamond marker represents the unitary stage duration $\tau = 0.001$ (sudden quench limit) and orange solid line with point marker represents that $\tau = 20$ (quasistatic case). The other parameters are the same with \textbf{Fig.~\ref{fig:Average work Vs Unitary process time}}.}
   \label{fig: Diagonal Probability Vs Unitary time G = 1 tau = 0.001}
\end{figure}

\begin{figure}[h!]
 \includegraphics[width=0.45\textwidth]{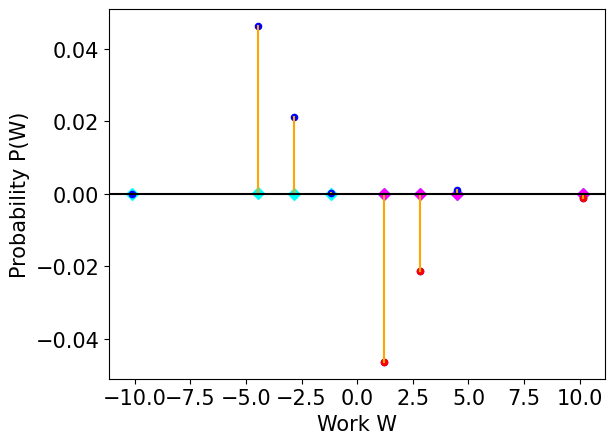}
   \caption{ Probabilities of the possible values of stochastic work originating from the off-diagonal contribution of the heat engine in a complete cycle for anisotropic interaction ($\gamma = 1$) between the spins. Point marker represents the unitary stage time $\tau = 0.001$ (sudden quench limit) and Diamond marker represents that $\tau = 20$ (quasistatic case). The other parameters are the same with \textbf{Fig.~\ref{fig:Average work Vs Unitary process time}}.}
   \label{fig:Off-diagonal Probability Vs Unitary time G = 1 tau = 0.001}
\end{figure}

The first term in the above expression arises due to the diagonal elements of the density matrix, 
which is the same as obtained by the TMP method, while the second term depends upon the off-diagonal elements. We plot this first term in \textbf{Fig.~\ref{fig: Diagonal Probability Vs Unitary time G = 1 tau = 0.001}} and the second term in \textbf{Fig.~\ref{fig:Off-diagonal Probability Vs Unitary time G = 1 tau = 0.001}} for $\gamma \ne 0$ and two different values of $\tau$. While the \textbf{Fig.~\ref{fig: Diagonal Probability Vs Unitary time G = 1 tau = 0.001}} displays usual positive values of the probabilities, there exist both positive and negative values of the probabilities in the \textbf{Fig.~\ref{fig:Off-diagonal Probability Vs Unitary time G = 1 tau = 0.001}}. It can also be observed that the sum of all values of these probabilities in the \textbf{Fig.~\ref{fig:Off-diagonal Probability Vs Unitary time G = 1 tau = 0.001}} becomes zero for any value of $\tau$, so that the total probability remains conserved. 

\subsection{Origin of the negative probability of stochastic work}

Now to determine the origin of the negative probability, we will conduct a detailed study of the work extraction stage $C \to D$, as this is a coherent process. From \textbf{Eq.~\ref{work stat w2}}, we have
\begin{equation}
    \begin{aligned}
        &P(W_2;\tau) = \rho^C_{00}(0) \abs{A^{CD}_{30}(\tau)}^2~ \delta\left\{W_2 - \left(\varepsilon^{D\tau}_3-\varepsilon^{C0}_0\right)\right\}\\ & + \rho^C_{33}(0)  \abs{A^{CD}_{33}(\tau)}^2~ \delta\left\{W_2 - \left(\varepsilon^{D\tau}_3-\varepsilon^{C0}_3\right)\right\}\\ &+ 2~\text{Re}[A^{CD}_{33}(\tau)\rho^C_{30}(0) A'^{CD}_{03}(\tau)]~ \\ &~~~~~~~~~~~~~~~~~~~~~~~~\times \delta\left\{W_2 - \left(\varepsilon^{D\tau}_3-\frac{\varepsilon^{C0}_0+\varepsilon^{C0}_3}{2}\right)\right\} \\ & + \rho^C_{00}(0)  \abs{A^{CD}_{00}(\tau)}^2 ~\delta\left\{W_2 - \left(\varepsilon^{D\tau}_0-\varepsilon^{C0}_0\right)\right\} \\ & + \rho^C_{33}(0) \abs{A^{CD}_{03}(\tau)}^2 ~\delta\left\{W_2 - \left(\varepsilon^{D\tau}_0-\varepsilon^{C0}_3\right)\right\}\\ & + 2~\text{Re}[A^{CD}_{03}(\tau)\rho^C_{30}(0) A'^{CD}_{00}(\tau)]~~ \\ &~~~~~~~~~~~~~~~~~~~~~~~~\times \delta\left\{W_2 - \left(\varepsilon^{D\tau}_0-\frac{\varepsilon^{C0}_0+\varepsilon^{C0}_3}{2}\right)\right\} \\ & + \rho^C_{11}(0) \abs{A^{CD}_{11}(\tau)}^2~ \delta\left\{W_2 - \left(\varepsilon^{D\tau}_1-\varepsilon^{C0}_1\right)\right\}\\ & + \rho^C_{22}(0) \abs{A^{CD}_{22}(\tau)}^2~ \delta\left\{W_2 - \left(\varepsilon^{D\tau}_2-\varepsilon^{C0}_2\right)\right\}
    \end{aligned}       
\end{equation}
In this distribution, the third and the sixth terms represent the interference between two probability amplitudes, i.e., between two paths of non-adiabatic transitions. As the probability amplitudes, $A^{CD}_{pq}(\tau) = A^{CD}_{03}(\tau), A^{CD}_{30}(\tau) $, can be negative, depending on the eigenstructures of the WS and the external control protocol, these interference terms can be negative. Consequently, the probabilities determined by these interference terms can be negative. If there are no off-diagonal terms, $\rho^C_{jk}(0) = \rho^C_{03}(0), \rho^C_{30}(0)$, in the density matrix at C, i.e., no coherence, then these interference terms become zero and we do not get any negative probability. %Therefore, we can say that quantum coherence allows us to get the interference effect, therefore, the negative probability 
Therefore, we can assert that quantum coherence enables the interference effect, which in turn results in negative probability \cite{solinas2022PRA, bednorz2010PRL}. These types of interference effects between two probability amplitudes in a measurement-based engine were indicated before \cite{purkait2023PRE}. %In the $A \to B$ work extraction process, as there is no quantum coherence, $\rho^A_{mn}(0) = 0$, in the initial density matrix at A, even though the probability amplitudes $A^{AB}_{pq}(\tau) = A^{AB}_{03}(\tau), A^{CD}_{30}(\tau)$ are negative, no negative probabilities will be observed. 
During the $A \to B$ stage, the absence of quantum coherence in the initial density matrix at A [i.e., $\rho^A_{mn}(0) = 0$] ensures that even if the probability amplitudes $A^{AB}_{pq}(\tau) = A^{AB}_{03}(\tau), A^{AB}_{30}(\tau)$ are negative, no negative probabilities will be observed.

\subsection{Average Work}
 % Check the diagonal and off-diagonal contribution

 % Show that only diagonal elements do not reproduce the same results as given by the density matrix method

%Now for a complete cycle, we can calculate the average value of work using the formula

\begin{figure}[h!]
\includegraphics[width=0.45\textwidth]{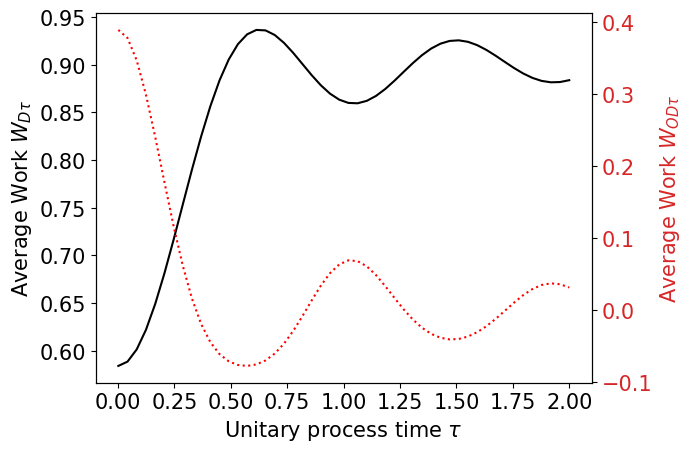}
   \caption{On the left axis (a) average work from the diagonal contribution of the probability distribution of work and on the right axis (b) average work from the off-diagonal contribution of the probability distribution of work as a function of unitary stage duration $\tau$. The other parameters are $\gamma = 1$ and the remaining are the same with \textbf{Fig.~\ref{fig:Average work Vs Unitary process time}}.}
   \label{Diagonal and off-diagonal contribution of average work}
\end{figure}

The average value of work in a complete cycle is given by
\begin{equation}\label{ave work}
    \langle W \rangle (\tau) = \sum_W P(W,Q_M;\tau) W\;.
\end{equation}
%where $w_n$ is all the possible stochastic works and $p_n(\tau)$ is the corresponding probability following the statistics of work \textbf{Eq.~\ref{work stat}}. 
Now if we plot this average work as a function of $\tau$ for different values of $\gamma$, we get the same plot as shown in \textbf{Fig.~\ref{fig:Average work Vs Unitary process time}}, obtained by the standard density matrix method (see \textbf{Sec.~\ref{cycle}}). %But, if we use the TPM method, the average work would not be the same as the standard method.
%To confirm that the FCS method is giving us the correct results, 
%\begin{figure}[h!]
% \includegraphics[width=0.4\textwidth]{Average work Vs Unitary process time using FCS.png}
%   \caption{Average work as a function of unitary process time ($\tau$). Other parameters are $\tau = 0.001$, $B_1 = 1$, $B_2 = 2$, and $T = 1$. }
%   \label{fig:Average work Vs Unitary process time by FCS}
%\end{figure}
%\subsection{Average work from the diagonal and off-diagonal terms}
%\subsubsection{Diagonal and off-diagonal contribution of average work}
We can express this average work (\textbf{Eq.~\ref{ave work}}) as combinations of the diagonal and off-diagonal contributions of the work distribution $\langle W \rangle (\tau) =\langle W_D \rangle (\tau) + \langle W_{OD} \rangle (\tau)$.
The plots of these two contributions as functions of $\tau$ are shown separately in \textbf{Fig.~\ref{Diagonal and off-diagonal contribution of average work}}.
In the short limit of $\tau$, the probabilities of certain values of stochastic work originating from the off-diagonal terms are non-zero (see \textbf{Fig.~\ref{fig:Off-diagonal Probability Vs Unitary time G = 1 tau = 0.001}}). This leads to $\langle W^{OD} \rangle (\tau) \neq 0$ (see \textbf{Fig.~\ref{Diagonal and off-diagonal contribution of average work}}), thereby increasing the total average work (see \textbf{Fig.~\ref{fig:Average work Vs Unitary process time}}). However, in the limit of very large $\tau$, these probabilities approach zero, resulting in $\langle W_{OD} \rangle (\tau) \simeq 0$. Therefore, for large $\tau$, the total average work consists only of the diagonal contribution, and the off-diagonal contribution is zero.

%These probability distributions suggest that the average works from all the probabilities and only from the diagonal terms should match in the long unitary process time. 

%\begin{figure}[h!]
% \includegraphics[width=0.4\textwidth]{Average work Vs Unitary time - off diagonal terms.png}
%   \caption{Probability as a function of work considering only the off-diagonal elements. Other parameters are $\tau = 20$, $\gamma = 1$, $B_1 = 1$, $B_2 = 2$ $T = 1$. }
%   \label{}
%\end{figure}

\begin{figure}[h!]
 \includegraphics[width=0.4\textwidth]{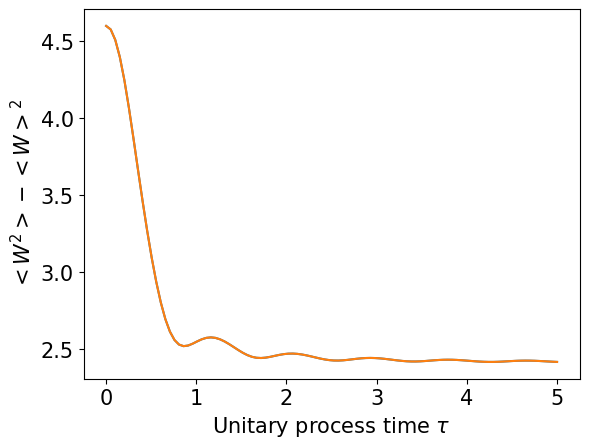}
   \caption{Fluctuation of work as a function of unitary stage duration $\tau$. The other parameters are $\gamma = 1$ and the remaining are the same with \textbf{Fig.~\ref{fig:Average work Vs Unitary process time}}. }
   \label{fig:work fluctuation Vs unitary time G = 1 tau = 0.001}
\end{figure}

{\it Work fluctuation:}
The fluctuation of work can be calculated as $\Delta W^2 (\tau) = \langle W^2\rangle (\tau) - \langle W\rangle^2 (\tau)$, where $\langle W^2\rangle (\tau) = \sum_W P(W,Q_M;\tau) W^2$. The plot of work fluctuation as a function of $\tau$ is shown in \textbf{Fig.~\ref{fig:work fluctuation Vs unitary time G = 1 tau = 0.001}}. It can be observed that in the small $\tau$ limit, the increased value of the average work is associated with enhanced work fluctuation.

% Fluctuation of work can be calculated as $\Delta W^2 (\tau) = \langle W^2\rangle (\tau) - \langle W\rangle^2 (\tau)$ , where $\langle W^2\rangle (\tau) = \sum_n w_n^2 p_n (\tau)$. The plot of the work fluctuation as a function of unitary process time is shown in \textbf{Fig.~\ref{fig:work fluctuation Vs unitary time G = 1 tau = 0.001}}. It can be seen that in the short limit of $\tau$, the increased average work is associated with increased fluctuation of work.

%Initially, at a very short instant of time, fluctuation of work is high, then it gradually reduces to a fixed value as unitary process time increases. 

%The nature of the work fluctuation for a measurement-based QOE is very similar to the work fluctuation of the QOE, which operates between two heat baths. 

%comment
%The plot of the work fluctuation for a QOE operates between two heat baths is shown in the \textbf{Fig.~\ref{fig:Work fluc Vs Unitary time for two heat baths by FCS G = 1 tau = 0.001.png}}.

%\begin{figure}[h!]
% \includegraphics[width=0.5\textwidth]{Work fluc Vs Unitary time for two baths by FCS G = 1 tau = 0.001.png}
%   \caption{Fluctuation of work as a function of unitary process time for a QOE operates between two heat baths. Other parameters are $\tau = 0.001$, $\gamma = 1$, $B_1 = 1$, $B_2 = 2$, $T_L = 1$, and $T_H = 10$. }
%   \label{fig:Work fluc Vs Unitary time for two heat baths by FCS G = 1 tau = 0.001.png}
%\end{figure}
%comment

\section{QOE operating with two heat baths}\label{Standard QOE}
\begin{figure}[h!]
\includegraphics[width=0.45\textwidth]{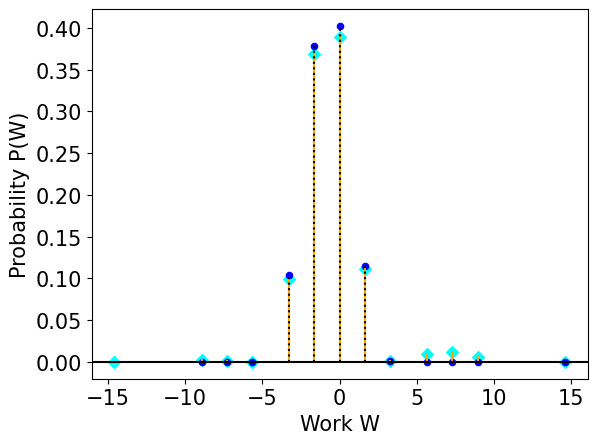}
   \caption{Probabilities of all possible values of stochastic work of a standard QOE, operating between two heat baths, in a complete cycle. The other parameters are $\tau = 0.001$, $\gamma = 1$, $T_H = 10$, $T_L = 1$ and the remaining are the same with \textbf{Fig.~\ref{fig:Average work Vs Unitary process time}}. }
   \label{fig:Prob Vs Work with two heat baths G = 1 tau = 0.001}
\end{figure}

Let us consider that a QOE is operating between two heat baths, where the isochoric heating stage $B \to C$ is carried out by coupling the WS with a heat bath \cite{purkait2024PRE}, instead of using non-selective quantum measurements, as discussed in \textbf{Sec.~\ref{QHE}}. We also consider the WS gets completely thermalized in both isochoric stages ($B \to C$ and $D \to A$). The probability distribution of work, which can be calculated by the FCS method (see \textbf{Eq.~\ref{off dia and dia con}}), is shown in \textbf{Fig.~\ref{fig:Prob Vs Work with two heat baths G = 1 tau = 0.001}}.  
We observe that there are no negative probabilities in the work distribution even for $\gamma \neq 0$ in the limit of $\tau \to 0$. This is because quantum coherence is absent at points A and C of the cycle. Therefore, we can conclude that the main difference between the work statistics in two types of QOEs (measurement-based and standard) is that the probability of stochastic work cannot be negative, at least for the complete thermalization case, in a standard QOE operating between two heat baths. In contrast, the quantum coherence can lead to negative probabilities of stochastic work in a measurement-based QOE. As the standard QOE operates between two heat baths, there is no initial quantum coherence with complete thermalization in the isochoric stages. %In this case, the FCS method and TPM method give the same distribution of work. 
Consequently, the FCS method and the TPM method yield the same distribution of work in this scenario.

\section{Conclusions}\label{Con}

Quantum non-selective measurements can be employed to fuel a QOE instead of using a heat bath, a configuration known as a measurement-based QOE. A previous study \cite{purkait2023PRE} demonstrated that a measurement-based QOE can perform better at finite times than the quasi-static limit in a coupled spin working system with an anisotropic interaction in between. To gain deeper insights into this phenomenon, we have investigated the work statistics of the same heat engine model.

The WS with an inter-spin anisotropic interaction exhibits quantum coherence in the energy eigenbasis at the beginning of a unitary work extraction stage. We found that the probability of certain values of stochastic work in the distribution can be negative with the presence of quantum coherence. Our analysis has revealed that quantum coherence induces interference terms in the probability distribution of work, and these interference terms can yield negative probabilities. Also, we found that coherence improves the average work in the faster unitary stage (i.e., for small $\tau$). 

Finally, we compared the work distribution of the measurement-based QOE with that of a standard QOE operating between two heat baths. We found that there is no negative probability of stochastic work in the distribution of a standard QOE due to the absence of quantum coherence when complete thermalization occurs in the isochoric stages.

%6. Importance of our results

\bibliography{ref}

\end{document}